\documentclass{camera-mod}
\usepackage{epsfig}

\def\beq{\begin{equation}}
\def\eeq{\end{equation}}
\newcommand\sss{\scriptscriptstyle\rm}
\newcommand\pt{p_{\sss {\rm T}}}
\newcommand\kt{k_{\sss {\rm T}}}
\newcommand\xt{x_{\sss {\rm T}}}
\newcommand\nE{n_{\sss E}}
\newcommand\as{\alpha_{\sss S}}

\begin{document}

%
\title{LHC PHYSICS: CHALLENGES\\FOR QCD}

%
\author{Stefano Frixione}
%
\organization{INFN, Sezione di Genova,
  Via Dodecaneso 33, 16146 Genova, Italy\\
  E-mail: {\em Stefano.Frixione@cern.ch}}

\maketitle

{\small
I review the status of the comparisons between a few measurements at
hadronic colliders and perturbative QCD predictions, which emphasize
the need for improving the current computations. Such improvements will
be mandatory for a satisfactory understanding of high-energy collisions 
at the LHC
}
\vskip 0.4truecm

One of the main goals of the LHC will be the search of the Higgs boson, the
only piece of the otherwise thoroughly tested Standard Model (SM) of which we
miss experimental evidence. Regardless of the existence of the Higgs boson,
LHC will likely shed light on the physics Beyond-the-SM (BSM), which should be
within reach, as LEP results and solar and atmospheric neutrino data
suggest. In this context, the role played by QCD may appear an ancillary
one. However, this seems to be quite at odd with the fact that LHC is a
hadronic collider, and strong interactions will be responsible for a prominent
part of the reactions taking place. On the other hand, one may claim that an
accurate knowledge of QCD predictions will not be necessary for the discovery
of -- say -- the Higgs in presence of a striking feature such as a narrow mass
peak, which could give the possibility of normalizing the background directly
with the data. However, if the discovery is based on a counting experiment, it
relies by definition on precise predictions for SM backgrounds, dominated
by QCD. These predictions will also prove essential for after-discovery
studies, when the properties of the newly-found particles will have to be
determined, or in the case of absence of BSM signals, in order to set limits
on BSM scenarios.

Having argued that QCD studies are one of the keys for a successful LHC
physics program, one question remains: are there still motivations to keep on 
working on them? The answer is again yes. The question is legitimate, since 
the striking success of perturbative QCD in predicting experimental results 
may lead to think that the current knowledge is sufficient to tackle the 
problems posed by the LHC. Unfortunately, this is not the case: because of
the large energy available, hard reactions will occur in a previously
unexplored regime, with peculiar kinematic characteristics, such as the
simultaneous presence of many, well-separated hard jets.  These new features
require to improve QCD predictions, either by increasing the accuracy of the
computations, or by considering issues which could have been safely neglected
up to now. Although one could give theoretical arguments for the improvements
that need to be achieved, it is also instructive to look at those measurements
which so far could not be described by perturbative QCD in an entirely
satisfactory way, which in fact, rather than hinting to a fundamental problem
of QCD, indicate that a deeper understanding is desirable of some aspects of
the computation. This is in fact relevant to the problem of QCD predictions
for the LHC, since aspects which may be marginal in the current 
phenomenological picture will be much more important in the future.

\begin{figure}[htb]
\begin{center}
\epsfig{file=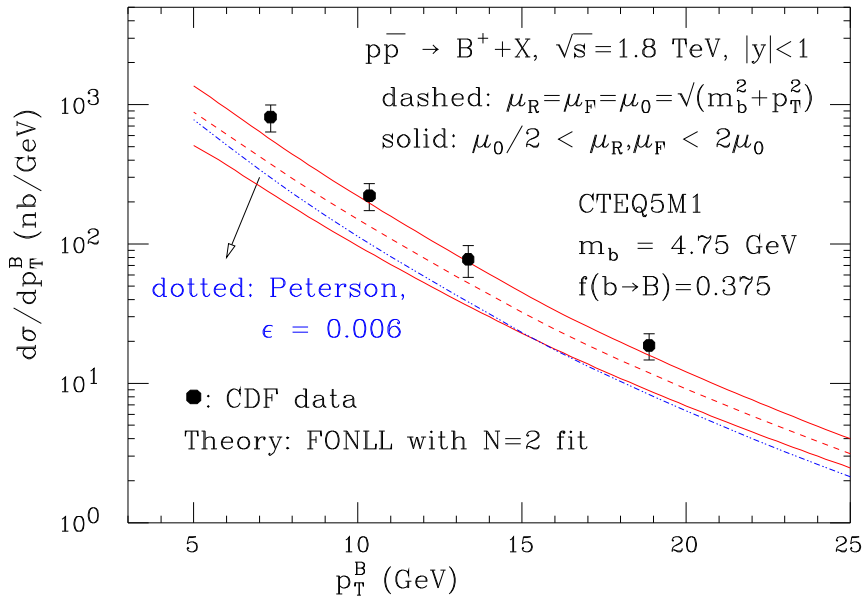,width=0.9\textwidth}
\end{center}
\caption{\label{fig:batcdf}
$B^+$ data~\cite{Acosta:2001rz} versus theoretical 
predictions~\cite{Cacciari:2002pa}.}
\end{figure}
Let me start with what has been seen historically as a major problem
in QCD, namely single-inclusive $b$ production at hadron colliders: 
although NLO QCD appears to give a good description of data in terms 
of shape, the rate is typically underestimated by a factor of 2. 
Recently, this picture seemed to receive strong support from a 
CDF measurement~\cite{Acosta:2001rz}, where the average of the data/theory 
ratio for $B^+$ mesons was quoted to be $2.9\pm 0.2\pm 0.4$. However,
it has been subsequently pointed out~\cite{Cacciari:2002pa} that
this value is mainly due to an improper treatment of the theoretical 
prediction, and that the correct result is $1.7\pm 0.5\pm 0.5$. This
value originates from using the state-of-the-art computation for 
single-inclusive $b$ $\pt$ spectrum (FONLL~\cite{Cacciari:1998it}),
and especially from the observation that the fragmentation function 
$b\to B$ as presently determined by using $e^+e^-$ data is not particularly
appropriate for the case of hadronic collisions. It is interesting to
note that the proper choice of the fragmentation parameters leads to
a comparison between theory and data (see fig.~\ref{fig:batcdf}) which
is basically identical to that relevant to $b$-jet transverse
energy~\cite{Frixione:1996nh,Abbott:2000iv}, i.e. to an observable 
independent of the details of the fragmentation mechanism. Therefore, $b$
production at hadron colliders should not be regarded any longer as a reason
of concern; on the other hand, the size of the theoretical uncertainties
(solid band in fig.~\ref{fig:batcdf}) prevents more stringent tests. It is
likely that a major source of improvement in this respect would be the
computation of the $b$ cross section to NNLO accuracy; a further enhancement
of the rate should be expected from
small-$x$~\cite{Catani:1990eg,Collins:1991ty} and threshold
resummations. Finally, let me remind that results for total rates for $b$
production in $ep$, $\gamma p$ and $\gamma\gamma$ collisions are in such a
disagreement with NLO QCD predictions, that no viable explanation for this
discrepancy has been found in any BSM scenario.  It is necessary to note that
in many cases the experimental results are extrapolated to the full phase
space from a rather narrow visible region.  However, it is encouraging that 
in a few cases the data are {\em also} presented without the extrapolation
outside the visible region, and in this way they are fully compatible with 
QCD predictions. This points out that some problem may be hidden in the Monte
Carlo (MC) simulation of heavy flavour production, especially in the low 
$\pt$ region.

Let me now turn to jet production at the Tevatron. The excess of CDF
single-inclusive jet data~\cite{Abe:1996wy} over NLO predictions at 
large $\pt$ has been regarded with much interest, being a potential
signal of new physics. On the other hand, it has been immediately 
observed~\cite{Huston:1995tw} that, by suitably adjusting the gluon 
density in the proton, one can obtain sets of PDFs (denoted with the
``HJ'' suffix by CTEQ) which give a decent global fit, and result in 
predictions for jets compatible with CDF measurements. This proves that
QCD has enough flexibility to accommodate the excess, but it is disturbing
that the ``HJ'' family is not the preferred one according to a (unweighted)
global fit procedure. This situation changes when D0 jet 
data~\cite{Abbott:2000kp} are included in the global fit, since the 
resulting PDF set turns out to belong to the ``HJ'' family (although it is 
{\em not} called accordingly). This happens because the Bjorken $x$'s relevant 
\begin{figure}[htb]
\begin{center}
\epsfig{file=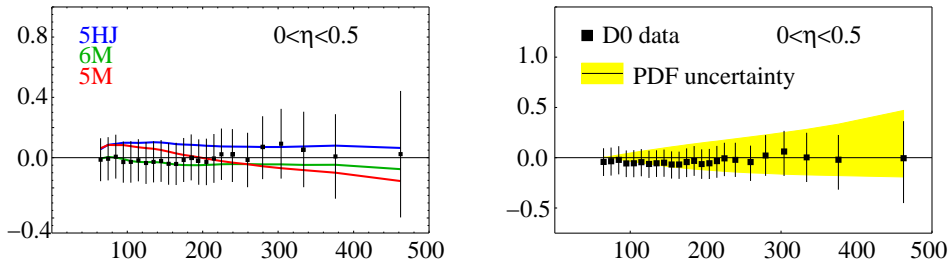,width=0.99\textwidth}
\end{center}
\caption{\label{fig:d0jets}
Data/theory ratios for single-inclusive jet production.
From ref.~\cite{Stump:2003yu}}
\end{figure}
to high-$\pt$ jet production are also relevant to small-$\pt$, large-$\eta$
production, a region probed by D0 (see ref.~\cite{Stump:2003yu} for
a discussion). The comparison between theory and data is now fully 
satisfactory: the left panel of fig.~\ref{fig:d0jets} presents the
ratio of the D0 data~\cite{Abbott:2000kp} over theoretical predictions
obtained with the PDF set CTEQ6M1~\cite{Stump:2003yu}; for reference, the
ratios of theoretical predictions obtained with other PDFs are also given;
the situation for CDF data is analogous. Although this excludes any evidence
of new physics in this channel, the situation may still change. This is
shown in the right panel of fig.~\ref{fig:d0jets}, where the band
represents the span of the theoretical predictions due to PDF uncertainties.
Thus, single-inclusive jet measurements at the Tevatron clearly
stress the importance of precise PDF determinations (which implies the
necessity of using data from many complementary processes), and of 
accurate estimates of the uncertainties affecting the PDFs. It is
also worth reminding that there are at least a couple of unpleasant 
aspects of jet production at colliders. Firstly, by reconstructing jets 
using a $\kt$-algorithm, D0~\cite{Abazov:2001hb} finds large discrepancies
with QCD predictions, a fact difficult to reconcile with the observation 
that, at the NLO and for suitable choices of jet-recombination parameters,
no major differences can be seen in the single-inclusive $\pt$ distribution
of jets reconstructed with the $\kt$ or the cone algorithms. Although
more experimental analyses with the $\kt$ algorithm are necessary in
order to confirm the result of ref.~\cite{Abazov:2001hb}, it is interesting 
to observe that hadronization corrections and the underlying event modelling,
as simulated by MC's, affect fairly differently the jets reconstructed with
different algorithms.  Secondly, the ratio of jet cross sections measured by
CDF and D0 at different c.m. energies ($\sqrt{S}=1.8$ and 0.63~TeV) and fixed
$\xt=2\pt/\sqrt{S}$, is not in agreement with QCD (which can predict this
quantity in a fairly accurate manner); besides, CDF and D0 data are also
mutually incompatible at low $\xt$. However, the differences are largely
within the uncertainties due to neglected power-suppressed effects. In both
cases, it appears that the experimental analyses would benefit from a much
deeper understanding of the interplay between perturbative and
non-perturbative physics, both at the level of MC's (whose use to estimate the
hadronization corrections applied to NLO predictions is rather empiric), and
in the context of approaches analogous to that of
DMW~\cite{Dokshitzer:1995zt,Dokshitzer:1995qm}.

\begin{figure}[htb]
\begin{center}
\epsfig{file=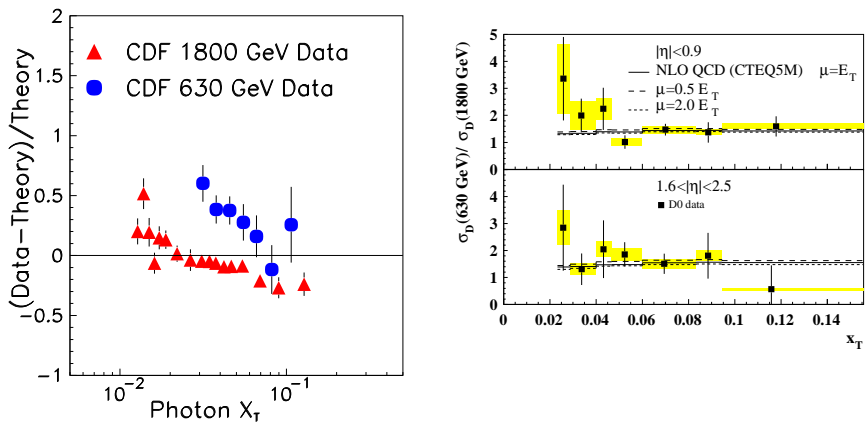,width=0.9\textwidth}
\end{center}
\caption{\label{fig:photons}
Comparison between prompt-photon data and QCD predictions}
\end{figure}
Let me finally deal with the case of $\gamma$ production. One of the reasons
of interest in this process is that, through the leading order partonic
reaction $qg\to\gamma q$, it can in principle constrain the gluon density
in a rather clean way. This is even more interesting if one observes that
the typical range in $\xt^{(\gamma)}$ probed by fixed-target experiments
is quite similar to the range in $\xt^{(jet)}$  probed in moderate- and
large-$\pt$ jet production at the Tevatron. Unfortunately, as discussed
in ref.~\cite{Aurenche:1998gv}, fixed-target photon data are not mutually
compatible. This nowadays prevents the use of these data in global PDF
fits. The problem is not peculiar to fixed-target experiments, since
the agreement between isolated-photon data at the Tevatron and QCD 
predictions has always been pretty marginal, as shown in the left panel
of fig.~\ref{fig:photons}, where CDF data~\cite{Acosta:2002ya} are
compared to NLO QCD results~\cite{Gluck:1994iz}. The situation, partly
because of larger statistical errors, improves for D0~\cite{Abazov:2001af}, 
as shown in the right panel of fig.~\ref{fig:photons} (here, ratios of cross
sections, rather than cross sections, are presented). It must be stressed 
that the isolation cuts used by experiments at colliders are unlikely to be
fully equivalent to those applied in theoretical computations. The whole
situation is rather inconclusive; if QCD predictions have to be used in the
context of analyses involving photons, it is urgent to devise a strategy that
allows one to apply the same cuts at the theoretical and experimental levels. 
For example, in a high-energy environment experiments prefer to define 
isolated photons with narrow cones; on the other hand, it is known that the
isolated-photon cross section at NLO becomes unphysical (being larger than the
fully inclusive one) for small isolation cone sizes~\cite{Catani:2002ny}. The
computation of yet higher orders, and especially of the all-order resummation
of the relevant terms, would be necessary in order to solve this problem.

Although no major problem for QCD emerges from the phenomenological
considerations given above, the picture would certainly look firmer
(not necessarily better) if we had: {\em a)} NNLO results\footnotemark[1]; 
{\em b)} resummed results, matched with fixed-order ones\footnotemark
\footnotetext{For certain processes and observables.}; {\em c)} better 
determinations of PDFs and their uncertainties; {\em d)} better understanding 
of the interplay between perturbative and non-perturbative physics; 
{\em e)} better models for the underlying event in MC's.
It should be clear that these issues are of primary importance
for LHC physics: large numbers of hard jets (and large K-factors), strong
impact of the accompanying non-hard events, and many-scale processes will 
occur plentiful. This is also worrying in view of the fact that MC's, the 
ubiquitous tools in experimental analyses, are not reliable when an
accurate description of large-angle emission is needed. It seems 
therefore mandatory to add to the list above: {\em f)} better MC simulation
of hard emissions.

Items {\em a)}--{\em f)} are all tough problems, but it is generally
believed, thanks to recent developments, that substantial progress
will be made before LHC comes into operation. In view of its relevance
to experimental collaborations, in the following I'll
concentrate on item {\em f)}. It is useful to briefly remind how an MC
works: for a given process, which at the LO receives contribution from 
$2\to n_0$ reactions, $(2+n_0)$-particle configurations are generated, 
according to exact tree-level matrix element (ME) computations. 
The quarks and gluons (partons henceforth) among these primary particles are
then allowed to emit more quarks and gluons, which are obtained from a parton
shower or dipole cascade approximation to QCD dynamics. To lessen the 
impact of this approximation on physical observables, one can devise
two strategies. The first aims at having $\nE$ extra hard partons in the final
state; thus, in the example given above, the number of final-state hard
particles would increase from $n_0$ to $n_0+\nE$. This approach is usually
referred to as {\em matrix element corrections}, since the MC must use the
$(2+n_0+\nE)$-particle ME's to generate the correct hard kinematics.  The
second strategy also aims at simulating the production of $n_0+\nE$ hard
particles, but improves the computation of rates as well, to N$^{\nE}$LO
accuracy. I'll generally denote the resulting MC as N$^{\nE}$LOwPS.

There are basically two major problems in the implementation of ME
corrections. The first problem is that of achieving a fast computation
of the ME's themselves for the largest possible $n_0+\nE$, and an
efficient phase-space generation. A variety of solutions exist nowadays
for this problem, implemented in packages which I'll denote as ME
generators. The second problem stems from the fact that multi-parton ME's are
IR divergent. Clearly, in hard-particle configurations IR divergences don't
appear; however, the definition of what hard means is, to a large extent,
arbitrary. In practice, hardness is achieved by imposing some cuts on suitable
partonic variables, such as $\pt$'s and $(\eta,\varphi)$-distances. 
I collectively denote these cuts by $\delta_{sep}$.
One assumes that $n$ hard partons will result (after the shower) into $n$
jets; but, with a probability depending on $\delta_{sep}$, a given $n$-jet
event could also result from $n+m$ hard partons. This means that, when
generating events at a fixed $n_0+\nE$ number of primary particles, physical
observables in general depend upon $\delta_{sep}$; I refer to this as the
$\delta_{sep}$-bias problem. Any solution to the $\delta_{sep}$-bias problem
implies a procedure to combine consistently ME's with different 
$n_0+\nE$'s. It should be stressed that, in presence of a
$\delta_{sep}$ bias, the interface of an ME generator (which is responsible
for producing the hard configurations, i.e. the initial conditions for
the shower), and a parton shower code is {\em not}, strictly speaking,
an event generator, since the events depend somehow on the value of
$\delta_{sep}$. In practice, the dependence is of the order of 20\%,
which is acceptable if one considers that, without ME corrections,
multi-jet configurations predicted by standard MC's are completely unreliable.
A solution to the $\delta_{sep}$-bias problem has been presented, for
$e^+e^-$ collisions, in ref.~\cite{Catani:2001cc} (CKKW henceforth), 
and subsequently extended (without formal proof) to hadronic collisions in 
ref.~\cite{Krauss:2002up}. Loosely speaking, CKKW achieve the following:
if an $n$-jet observable is affected by the $\delta_{sep}$ bias in
the following way
\beq
\sigma_n\sim \as^{n-2}\sum_k a_k\as^k\log^{2k}\delta_{sep}\,,
\eeq
by applying the CKKW prescription one gets
\beq
\as^{n-2}(\delta_{sep}^a+
\sum_k b_k\as^k\log^{2k-2}\delta_{sep})\,.
\eeq
The implementation of CKKW in popular event generators for hadronic
collisions in under way, and it will have reached a mature and
well-tested stage when LHC will come into operation.

The implementation of an N$^{\nE}$LOwPS can be seen as an upgrade of ME
corrections: not only one wants to describe the kinematics of $n_0+\nE$ hard
particles correctly, but the information on N$^{\nE}$LO rates must also be
included. First attempts at solving this problem have only recently become
available, and only for the case $\nE=1$. The striking feature of an NLOwPS
is the computation of loop diagrams (which are necessary in order to
compute total rates to NLO accuracy); this in general implies the presence
of negative weights. This is a new feature in MC's, which however doesn't
spoil their probabilistic nature. In fact, in NLOwPS the distributions 
of positive and negative weights are {\em separately} finite, at variance
with what happens in NLO computations; thus, each of them can be unweighted
and evolved separately, since no cancellation between large numbers is
involved in this procedure. On the other hand, the contribution of loop 
diagrams implies that the $\delta_{sep}$-bias problem which affects ME
corrections is simply not present. At the moment, the following codes
implement different prescriptions for NLOwPS in hadronic collisions:
$\Phi$-veto~\cite{Dobbs:2001dq},
MC@NLO~\cite{Frixione:2002ik,Frixione:2003ei},
GRACE\_LLsub~\cite{Kurihara:2002ne}.
$\Phi$-veto is based on the slicing method, and features $Z^*$ and $W^*$
production; it is affected by double counting according to the definition
of ref.~\cite{Frixione:2002ik}, but numerically this problem seems to
be of minor importance; it is interfaced with Herwig and Pythia. MC@NLO 
is based on the subtraction method, and features $W^+W^-$, $ZZ$, $WZ$, 
$b\bar{b}$, $t\bar{t}$, SM Higgs, $Z$, $W$, and $\gamma$ production; it
is interfaced with Herwig. GRACE\_LLsub is based on the slicing method 
and on the fully-numerical computation of the matrix elements;
it features Drell-Yan production, and is not affected by double counting
only if the parton shower of ref.~\cite{Kurihara:2002ne} is adopted
(i.e., other showering codes cannot be used at the moment). The field
of NLOwPS, still behind that of ME corrections, is rapidly evolving,
and more ideas will appear in the future; soon, more processes will 
be implemented, and a thorough comparison between the various approaches 
will have to be made.

In summary, a lot of interesting developments are currently occurring in QCD,
which will provide a solid benchmark for LHC studies. It will be vital for
experimental collaborations to exploit these results, both by using new MC
tools (with ME corrections and NLOwPS) in the course of their analyses, and by
considering the most precise theoretical results available, at fixed-order
(NNLO) or in resummed computations (with NLL or NNLL accuracy). In the coming
years, it will also be crucial to learn from the experience of HERA and the
Tevatron, which will hopefully provide the necessary data to determine PDFs at
an unprecedented level of accuracy, and to test the various models for
underlying events.

It's a pleasure to thank the organizers for an interesting meeting;
as a phenomenologist, I think it is beneficial for both theoretical
and experimental communities to have frequent (elastic) interactions.

\end{document}